# Influence of a carbon over-coat on the X-ray reflectance of XEUS mirrors

D H Lumb[a], F E Christensen[b], C P Jensen[b], M Krumrey[c]

a) Advanced Concepts and Science Payloads Office, European Space Agency, ESTEC, Postbus 299, 2200AG Noordwijk, Netherlands

b) Danish National Space Centre, DK-2100 Copenhagen Ø Juliane Maries Vej 30, Denmark

c) Physikalisch-Technische Bundesanstalt, Abbestr. 2-12, 10587 Berlin, Germany

**Abstract**

We describe measurements of the X-ray reflectance in the range 2 to 10 keV of samples representative of coated silicon wafers that are proposed for the fabrication of the XEUS (X-ray Evolving Universe Spectrometer) mission. We compare the reflectance of silicon samples coated with bare Pt, with that for samples with an additional 10nm thick carbon over-coating. We demonstrate a significant improvement in reflectance in the energy range ~1 to 4 keV, and at a grazing incidence angle of 10 mrad (0.57°). We consider the resulting effective area that could be attained with an optimized design of the XEUS telescope. Typically an improvement of 10 to 60 % in effective area, depending on photon energy, can be achieved using the carbon overcoat.

## 1. Introduction

Grazing incidence X-ray telescopes have been used for decades as a core element of space-borne X-ray astrophysics observatories, most recently in the case of the XMM Newton[1] and Chandra[2] telescopes. Typically they comprise Wolter 1 hyperbola-parabola mirror pairs, arranged in a highly nested co-axial configuration. High reflectance up to photon energies of a few keV is secured through the use of highly polished, high density, high atomic weight coatings, most often gold.

During the calibration campaigns for the Chandra mirror effective areas, it was determined that the optical constants for the Ir reflecting layer could best be reconciled by considering a small amount of carbon contamination on the mirror surface[3]. A surprising discovery was the modest increase in reflectance in the energy range 2 to 5 keV, but the concern at the time was mainly to limit *changes* in the film thickness so as to avoid an uncontrolled modification in effective area between ground calibration and in-flight measurements.

Subsequently, Pareschi et al [4] pointed out that the deliberate employment of a low density external film over any mirror acts to reduce the photoelectric absorption when the mirror is in the total external reflection regime. In such a case the effective area of the mirror could intentionally be increased between 1 and 4 keV.

Previous activities have been described in the context of developing graded depth multi-layer coatings to enhance the hard X-ray reflectance for the XEUS mission [5]. With evolving mission configuration and science priorities [6,7], the emphasis for maximizing the potential for XEUS performance has been modified. The science requirements now request the maximization of effective area at ~ 1 keV, in order to optimize sensitivity near the peak of X-ray photon flux in targets, especially where cosmological red shift tends to bias spectra to the soft X-ray band.

We report here on an investigation into the effect of a deliberately introduced carbon layer over a platinum reflecting surface. We present measurements of the reflectance vs. graze angle at selected energies, and of an energy scan for the reflectance at a typical graze angle for the XEUS mirrors. The data are compared with models and used to verify predictions for the beneficial effect on the XEUS effective area

## 2. XEUS

Based on a novel optics technology a new generation of X-ray space telescope design is being investigated by the European Space Agency, in preparation for potential consideration in the future Cosmic Visions 2015-2025 Science Programme. The core of the telescope concept is the Silicon based High resolution Pore Optics (HPO) [8,9]

Details of the production and assembly process are provided in reference [10]. The X-HPO units are fabricated starting from high quality selected 12" (300 mm) Si wafers, polished with a flatness better than 180 nm (peak-to-valley, measured over 25x25 mm$^2$) and equipped with ribs on the back side. Processed silicon wafer components are then stacked onto a precision Si mandrel, requiring only a single curvature (for a conic surface that approximates one of the Wolter paraboloid/hyperboloid pairs). Several plates are stacked on top of each other while being curved in the azimuthal direction to form a single monolithic unit that is intrinsically very stiff, as well as possessing very good temperature stability without differential expansion problems

The production of these mirrors is effectively mastered through the huge effort invested by the semiconductor industry into the production of highly polished substrates, and the development of very well established production procedures. The demonstration of the mounting of the plates into units easy to handle, integrate and align is also well advanced. Two HPO units, as required for the implementation of conically approximated Wolter-I elements, are joined into tandems by the use of dedicated CeSiC ceramic structures. This assembly requires very high precision and is therefore, in the on-going R&D phase, performed at a synchrotron radiation facility at a dedicated beamline in the laboratory of the Physikalisch-Technische Bundesanstalt (PTB) at the storage ring BESSY. This complements and verifies the optical and mechanical metrology systems used in the integration system. Over small test article dimensions, the required performance of 5 arcseconds resolution is already being approached. A representative azimuthal "petal" section comprising many individual HPO modules is being fabricated and furnished with test modules. The petal contains a number of identical HPOs covering one radius, and with typically 15 to 20 different radial sections with different HPO configuration. From this activity an engineering model representative of the XEUS telescope configuration will be achieved.

A crucial part of the telescope design is the definition of the mirror effective area as a function of the photon energy. This task is particularly important as it impacts not only on science but also very significantly on the overall system in terms of total spacecraft (S/C) mass, formation flying requirements, baffling design and configuration. Although iterations are to be expected as the project evolves, a preliminary analysis has been conducted.

The analysis performed to date is based on the assumption of a nominal focal length of 35 m and a geometric area available to the mirror limited by the launcher fairing and by the inner S/C body (radii between 0.66 and 2.2 m). Different assumptions have been made on the adoption of possible single layer coatings on the HPO units, such as bare silicon, uniform iridium, gold and nickel coatings (see Figure 1). For each silicon plate in the assembly, the geometric projected frontal collecting area is calculated, and for the on-axis rays the grazing angle calculated and used to determine the energy dependent reflectance for that location. The total on-axis effective area is then obtained by summing over all plate radii for the product of collecting area and reflectance (accounting for two reflections).

Two features are immediately apparent:
- achievement of a required >5 m$^2$ at 1 keV may not be attained while satisfying further effective area requirements at other specified energies (e.g. > 2 m² at 6 keV) of scientific interest because the energy response of different metal coatings are markedly different,
- and the M-edge absorption features around 1 - 2 keV are very marked and would lead to calibration and data analysis challenges.

*Figure 1 Comparison of achievable effective area for XEUS optics when applying different monolithic reflective coatings. The effective area in each case around 1keV is similar, however the overall energy response and depth of absorption edges is markedly different. A heavy material is needed for high reflectance at energies up to 10 keV, but these materials have M-edges around 2 keV, resulting in pronounced variations and a generally lower reflectance just above the edges.*

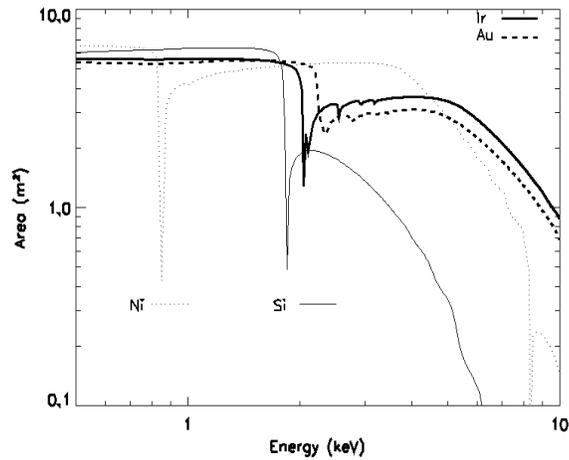

### 3. Deposition

As part of the on-going programme of developing graded depth multi-layer coatings for XEUS, some effort has already been expended in the deposition of Pt and C bi-layers, which have been attempted on silicon strips that were cut from the nominal XEUS pore optics substrate materials. It was therefore decided to use Pt as a reference monolithic metal coating, and compare the achieved reflectance with that obtained from similar samples coated with a carbon final layer. A number of samples were prepared, but Table 1 lists the nominal layer thicknesses for the samples subsequently measured at the test facility. The coating was performed in a planar dc magnetron sputtering facility of the Danish National Space Center [11].

### 4. Measurements

The samples were measured using the Four-Crystal Monochromator (FCM) of the PTB laboratory at BESSY. The FCM beamline [12] covers a photon energy range of 1.75 keV to 10 keV. The radiation is focused with a toroidal mirror and a plane mirror which can be cylindrically bent. Two different monochromator crystal sets, consisting of either four Si(111) or four InSb(111) crystals, can be interchanged in vacuum. The four-crystal configuration assures already high spectral purity, but the higher order content for nominal photon energies below 4 keV is further reduced by selecting a $MgF_2$ coating stripe on the bendable mirror instead of the Pt coating. The sum of all higher order contributions to the total photon flux remains below $3 \cdot 10^{-4}$ at all energies and even below $3 \cdot 10^{-5}$ above 2.5 keV, decreasing further towards higher energies. . The mirrors were placed in a UHV reflectometer that provides 0.001° angular resolution for sample and detector [13]. Silicon photodiodes with different apertures and a counting detector are mounted on the detector (2θ) arm. An additional thin photodiode operated in transmission in front of the reflectometer is used for normalization. For the measurements presented here, the reflectance was determined as the ratio of the normalised current for one of these diodes in the reflected beam to its normalized current in the direct beam.

*4.1. θ/2θ Scans*

θ/2θ scans were performed at two energies; at 8.048 keV to facilitate comparison with potential measurements in laboratory reflectometers, and at 2.8 keV that is near the peak energy in expected reflectance improvement.

*Figure 2  θ/2θ scan at a fixed energy of  8 keV for the samples Xeusb20 (- - - - ) and Xeusb21 (———). The periodicity in reflectance are primarily determined by the thickness of Pt layer, which in these samples differ by ~ 5%*

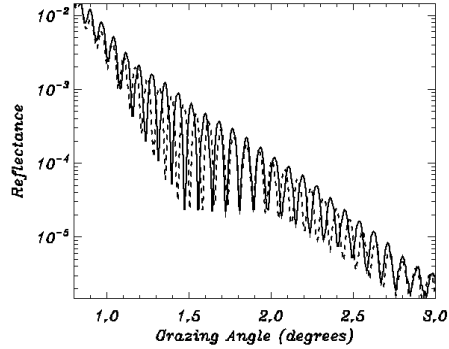

At 8 keV the difference in response between the 3 samples was not significant (Figure 2). At 2.8 keV it was expected that the maximum difference in reflectance would be manifested. Figure 3 compares the measured data with best fit modeled response according to the parameters in table 2.

For the XEUS telescopes, with the mirror radii in the range 0.66 to 2.2 m and a focal length 35m, the grazing incidence angle range is 0.3 ° to 0.9 °, therefore from Figure 3 it can be seen that the reflectance with C overcoat is significantly higher than for bare Pt for all angles of interest

*Figure 3 θ/2θ scan at 2.8 keV;*  **X**  *Xeusb20 (no C overcoat)*  **+**  *Xeusb21 (10 nm C overcoat ) (In each case the full line is the model based on the best fit parameters)*

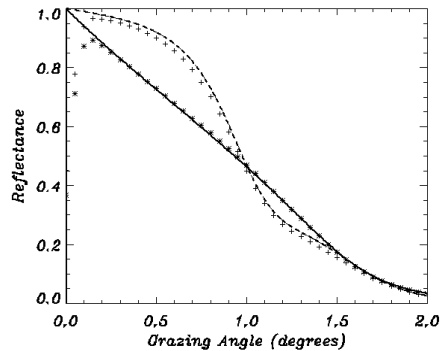

*4.2. Energy Scans*

For a typical grazing angle chosen for a mirror radius in the middle of the XEUS telescope range (0.57°), we performed an energy scan to confirm the energy dependence of reflectance improvement. This is shown in Figure 4, where the reflectance was calculated using Fresnel equations modified by the Nevot & Croce description of interface roughness imperfections. The calculations were implemented by the IMD code which allows a convenient fitting minimization using the Marquardt algorithm. The details of implementation are provided in [14]. The best match between theory and measured data is made assuming the parameters in Table 2. The Pt surface roughness compares with ~0.5 nm (r.m.s.) achieved with interface layers of graded depth multilayers [5]. The Si layer roughness is somewhat larger than normally achieved, and the C layer roughness was not formally distinguishable from that of the Pt, so was simply tied to the former. A best fit was achieved assuming an apparent density of Pt a little lower than the nominal. The fit was insensitive to density of carbon. For the Xeusb20 sample a thin layer of C is also required formally, and this we assume to represent a monolayer of contamination.

Figure 4 Energy scan at grazing angle 0.57°. x Xeusb20 (no C overcoat). + Xeusb21 (10 nm C overcoat). The full lines represent the best model based on parameters determined from angles scans.

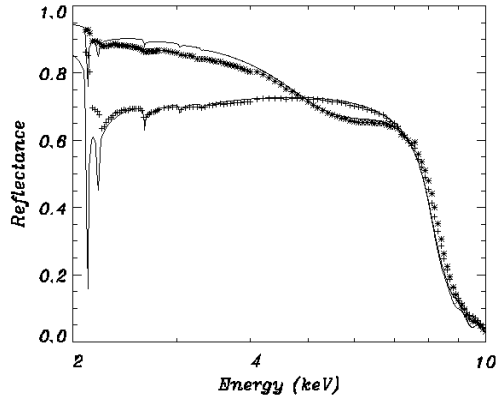

**5. XEUS baseline performance**

As reported in Lumb et al [5], there remain challenges for the transfer of the multi-layer coating process to the industrial process for silicon wafer stacking necessary for high quality mirror optic production. In particular the compatibility of coatings with the adhesion properties required of the rib placement on high quality surface finish needs to be confirmed. Such activity is already under investigation, and notwithstanding the practical implementation details, we use the promising results reported above to extrapolate to an eventual improvement in XEUS effective area performance. The smoothing of M-edge absorption edge features allows for very similar effective area for all feasible metallic coatings, and therefore we choose to optimize for the highest effective area at energies approaching 10 keV.

Figure 5 thus demonstrates the effective area of the baseline XEUS telescope design using either a bare Ir coating or one with Ir coated with carbon. The thickness of this carbon has been modeled as smoothly increasing from 5 nm to 15 nm in progressing from the outermost to the innermost mirror radii.

In order preserve the good angular resolution properties of the telescope, the figure and surface roughness of the silicon plates must be maintained. In this work the roughness of the top layers exceeded by a factor ~2 the required roughness and that which has already been demonstrated on the associated graded-depth multilayer coatings. The source of this degradation has not yet been identified. With the small sample strips employed to date, flatness sufficient to conclude the measurements was achieved by employing a vacuum chuck. The small size was dictated by availability and the deposition chamber geometries, and future activities will need to meet the flatness specification after coating. However the calculations of coating induced stress suggest this is lower than any changes due to thermal stresses between coating environment and in-orbit (cooled) conditions, and hence not likely to be a fundamental problem

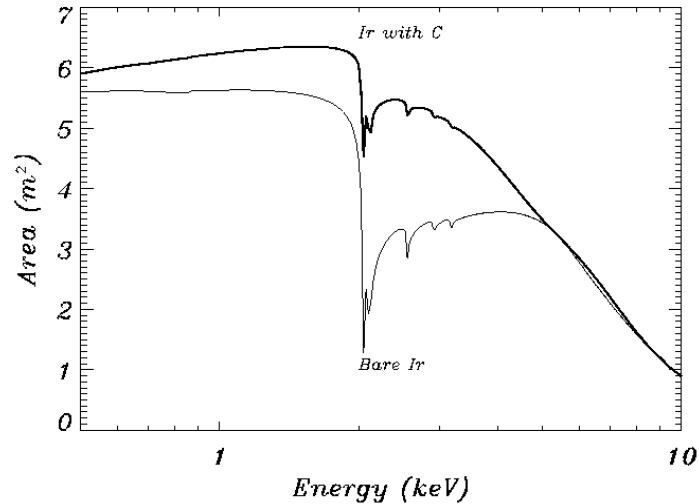

*Figure 5 Estimate of the expected collecting area of the XEUS telescope when using either bare Ir or C overcoated Ir as the reflecting surface. An increase of 10% and 35% is estimated at 1 and 3 keV respectively*

## 6. Conclusions

The telescope for the XEUS mission will make use of novel fabrication technology that is based on the advances in silicon wafer fabrication developed for the microelectronics industry. To achieve high reflectance at X-ray energies requires very smooth surfaces coated with material of high density and atomic weight. We have demonstrated that the silicon samples representative of XEUS technology can be deposited with Pt with low roughness, and then coated with a thin carbon layer to further enhance the reflectance.

The measured reflectance data are well fit with layer properties that are close to those of the nominal deposited depths. The addition of the carbon coating significantly increases the reflectance at a typical graze angle for X-ray telescopes and at energies in the range 1 to 5 keV as predicted. The consequent improvement of effective area for the specific case of the XEUS baseline telescope design varies from 10 to 60% depending on energy within the range 0.8 to 4.5keV.

## 7. Acknowledgements

We acknowledge gratefully the assistance of Levent Cibik (PTB) in conducting the measurements at BESSY. The XEUS Science Advisory Group are thanked for their recommendations on scientific requirements for a future X-ray astronomy mission, the XEUS Telescope Working Group, especially G Pareschi for advice on XEUS telescope design, and N Rando at ESTEC for details on the physical envelope available with different launch vehicles.

| Sample | Pt thickness / nm | C thickness / nm |
|---|---|---|
| Xeusb20 | 53.5 | 0 |
| Xeusb22 | 49.2 | 10 |
| Xeusb23 | 53.6 | 10 |

**Table 1** Three samples were measured using synchrotron radiation. Their nominal layer thicknesses were determined by 8 keV reflectometry after deposition at DNSC

|  | Xeusb20 | Xeusb23 | Xeusb21 |
|---|---|---|---|
| Pt thickness / nm | 53.3 ± 0.5 | 53.0 ± 0.6 | 49.1 ± 2.0 |
| Pt Surface roughness / nm | 0.7 ± 0.1 | 0.6 ± 0.1 | 0.7 ± 0.1 |
| Pt relative density | 98% | 98% | 95% |
| Si surface roughness/ nm | 0.3 ± 0.1 | 0.3 ± 0.1 | 0.3 ± 0.1 |
| C thickness/ nm | 0.3 ± 0.2 | 9.7 ± 0.2 | 9.7 ± 0.8 |

Table 2 Properties of the deposited layers of three samples that are derived from a best fit to the reflectance data sets ($\theta/2\theta$ and energy scans) for each sample.